\documentclass[runningheads]{llncs}

\usepackage[T1]{fontenc}
\usepackage{cite}
\usepackage{hyperref}
\usepackage{bm}
\usepackage{multirow}
\usepackage{booktabs}
\usepackage[caption=false]{subfig}
\usepackage{amsmath,amssymb,amsfonts}
\usepackage{algorithmic}
\usepackage{graphicx}
\usepackage{textcomp}
\usepackage{xcolor}
\usepackage[misc]{ifsym}

\begin{document}

\title{Cross-Domain Causal Preference Learning for Out-of-Distribution Recommendation}
\titlerunning{CDCOR}

\author{Zhuhang Li \and Ning Yang$^{(\textrm{\Letter})}$}
\authorrunning{Z. Li and N. Yang}

\institute{School of Computer Science, Sichuan University, Chengdu, China
\email{lizhuhang@stu.scu.edu.cn}\\
\email{yangning@scu.edu.cn}}

\maketitle

\begin{abstract}
Recommender systems use users' historical interactions to learn their preferences and deliver personalized recommendations from a vast array of candidate items. Current recommender systems primarily rely on the assumption that the training and testing datasets have identical distributions, which may not hold true in reality.
In fact, the distribution shift between training and testing datasets often occurs as a result of the evolution of user attributes, which degrades the performance of the conventional recommender systems because they fail in Out-of-Distribution (OOD) generalization, particularly in situations of data sparsity.
This study delves deeply into the challenge of OOD generalization and proposes a novel model called Cross-Domain Causal Preference Learning for Out-of-Distribution Recommendation (CDCOR), which involves employing a domain adversarial network to uncover users' domain-shared preferences and utilizing a causal structure learner to capture causal invariance to deal with the OOD problem. Through extensive experiments on two real-world datasets, we validate the remarkable performance of our model in handling diverse scenarios of data sparsity and out-of-distribution environments. Furthermore, our approach surpasses the benchmark models, showcasing outstanding capabilities in out-of-distribution generalization. The code and datasets are available at: \url{https://github.com/Rexhaha/CDCOR}.

\end{abstract}

\section{Introduction}
\begin{figure}
\centering
\includegraphics[width=0.7\textwidth]{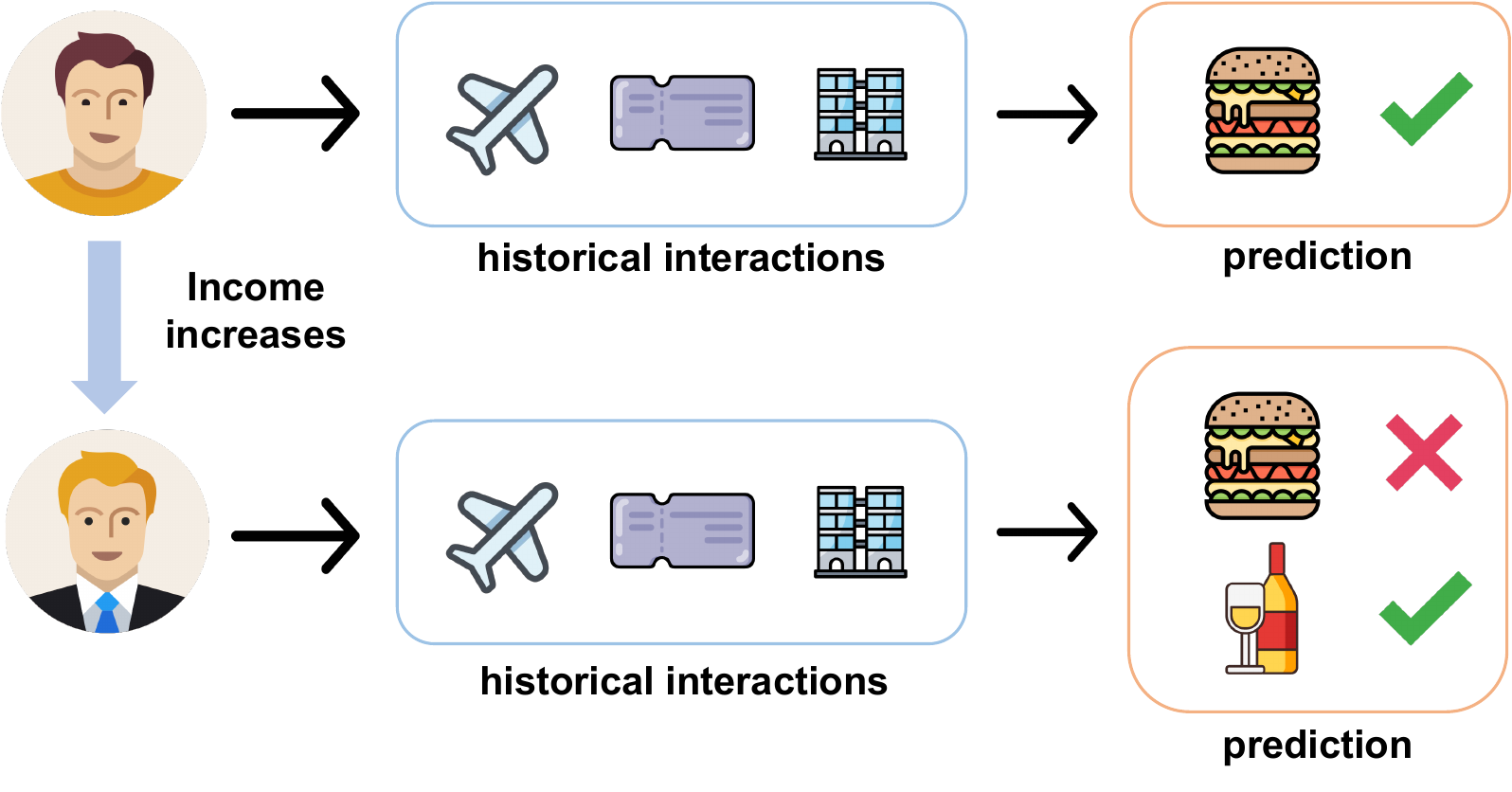}
\caption{Example of OOD recommendation.}
\label{fig1}
\end{figure}

Recommender systems aim to alleviate the problem of information overloading with the personalized preference learned from the historical interactions of users.
Existing approaches commonly assume an identical distribution during training and testing phases, which may not hold true in reality.
In fact, the distribution of user preferences frequently shifts over time as result of the changes of user attributes.
For example, the food preference of the user while traveling shown in Fig.~\ref{fig1} changes from burgers to wine due to increased income.
The user preferences learned from outdated interaction data and user attributes can lead to inaccurate recommendations, which can negatively impact the user experience.
Consequently, improving recommender systems performance in Out-of-Distribution (OOD) environments has emerged as a pressing problem~\cite{liu2021towards}.

There has been limited research on OOD generalization in recommender systems.
Some existing approaches address the OOD problem in recommender systems as a debiasing issue. 
DICE~\cite{zheng2021disentangling} divides the factors of users interact with items into two parts: interest and conformity, and makes each embedding capture only one cause by training with cause-specific data.
DCCL~\cite{zhao2023disentangled} incorporates item popularity as a weight within the InfoNCE to mitigate popularity bias.
While these methods effectively address bias, they do not explicitly consider situations that are more common in the real world, such as changes in user attributes.

Some approaches tackle the issue of changes in user attributes by employing a causal inference perspective~\cite{wang2022causal,wang2023causal,he2022causpref}.
Since the causal structure reflecting the user preferences generation process can be kept invariant in the data distribution shift, recommender systems can obtain the ability to capture the user preferences generation process which is independent of the data distribution by learning causal structure.
COR~\cite{wang2022causal} considers the changes of user attributes as an intervention, and OOD recommendations as the probabilistic inference of interactions following the intervention. However, this approach relies on an artificial designation of causality, which leads to a strong correlation between the performance of the model and the prior knowledge of the expert.
CausPref~\cite{he2022causpref} learns the causal structure from the data via a directed acyclic constraint, which extends the use of causal inference-based recommender systems. However, challenges of learning causal structure from data still remain:
\begin{itemize}
\item \textbf{Data sparsity}: data sparsity makes recommender systems difficult to learn the correct causal structure. On the one hand, data sparsity increases the risk of model overfitting, and on the other hand, data sparsity may lead to confusion between causality and correlation.
\item \textbf{Explicit attributes are difficult to obtain}: learning causal structure from data requires the dataset to contain explicit attributes of user and item. 
However, obtaining such attributes can be problematic due to privacy restrictions imposed by individual platforms, which leads to the failure of the method.
Even though we could encode the user IDs as latent attributes, this is affected by outdated interactions and requires dense data to encode correctly.
\end{itemize}

To tackle these problems, we propose a novel cross-domain recommendation model called \textbf{C}ross-\textbf{D}omain \textbf{C}ausal Preference Learning for \textbf{O}ut-of-Distribution \textbf{R}ecommendation (CDCOR).
The main idea of CDCOR is to utilize the data-rich source domain to help the model to learn the causal relationship between user attributes and user preferences in data-sparse target domain, which ensures that the user latent attributes be encoded correctly.
In this paper, we assume that the source and target domains share users, which is the most common scenario in cross-domain recommendation. These users act as a bridge between the two domains, allowing knowledge from the source domain transfer to the target domain in order to help the model learns the causal structure and encodes correct user latent attributes in target domain.
However, learning causal structrue from different domains is not easy due to the following challenges: 1) discovering the common parts of causality across different domains poses significant obstacles; 2) effectively utilizing the common parts of causality to improve the OOD recommendation performance of the model in target domain presents another challenge.

To address the above challenges, we design a domain-shared preference encoder to extract user's domain-shared preferences from their latent arrtibutes. Through an adversarial training process of the domain-shared preference encoder with the domain discriminator, the property that domain-shared preferences are independent of the domain is guaranteed.
Then we use a Directed Acyclic Graph (DAG) learner to model causal structure.
The causal structure learned from domain-shared preferences is also domain-shared, so that the knowledge of the source domain can be used to help improve the OOD recommendation performance of the model in the target domain.
To ensure the distinctiveness of user preferences across domains, CDCOR incorporates a user domain-specific preference encoder in each domain, allowing for the capturing of domain-specific preferences.
The main contributions of this paper can be summarized as follows:
\begin{itemize}
\item We propose a novel model called Cross-Domain Causal Preference Learning for Out-of-Distribution Recommendation (CDCOR), which uses source domain knowledge to help the model to improve their OOD recommendation performance in target domain.
To our best knowledge, this work is the first recommendation model that uses cross-domain knowledge to solve the OOD problem.
\item We extend causal structure learning from explicit attributes to latent attributes, enriching the usage scenarios of causal inference-based recommender systems.
\item We conduct extensive experiments to demonstrate the effectiveness of our approach in dealing with various OOD scenarios.
\end{itemize}

\section{Preliminaries}
\subsubsection{Problem Definition.}
In this paper, we assume that the input data take form of implicit feedback such as click records. 
We consider two domains, a source domain $D^s$ which has rich data, and a target doamin $D^t$ which only has few data. 
The users are denoted as $U=\{u_1,u_2,...,u_m\}$. 
The item sets from the source and target domains are $I^s=\{i^s_1,i^s_2,...,i^s_{n^s}\}$ and $I^t=\{i^t_1,i^t_2,...,i^t_{n^t}\}$ respectively, with the corresponding interaction matrices given by $\bm{Y}^s$ and $\bm{Y}^t$. 
Each element $y_{ui}^s\in\{0,1\}$ and $y_{ui}^t\in\{0,1\}$ in these two matrices indicates whether there is an interaction between user $u$ and item $i$. 
Then the target task is the single-target recommendation, where we predict $\hat{y}_{ui}^t(u\in U,i\in I^t)$, via utilizing the knowledge from the $\bm{Y}^s$ and $\bm{Y}^t$.
\subsubsection{Out-of-Distribution Recommendation.}
Given the training dataset $D_{tr}$ and testing dataset $D_{te}$, where the samples satisfy the training distribution $P_{tr}(U,I)$ and testing distribution $P_{te}(U,I)$ respectively. In OOD recommendation, the testing distribution may deviate from training distribution, that is $P_{tr}(U,I)\ne P_{te}(U,I)$. The goal of OOD recommendation is to train a recommendation model from training distribution $P_{tr}(U,I)$ to predict the probability of user-item interactions in the testing distribution $P_{te}(U,I)$.

\section{Proposed Method}
\renewcommand\floatpagefraction{.9}
\begin{figure}
\centering
\includegraphics[width=\textwidth]{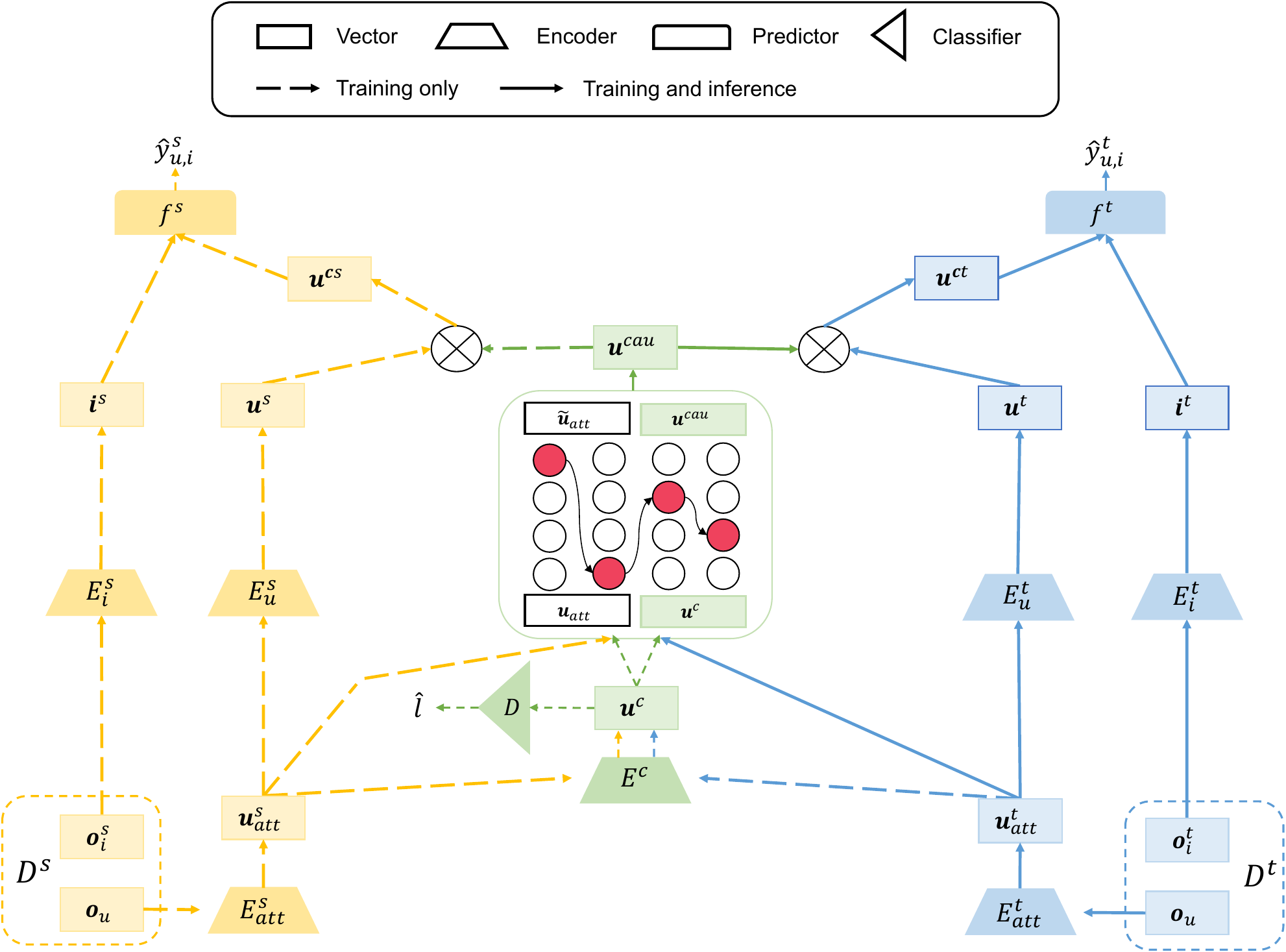}
\caption{The Architecture of CDCOR. The source domain is in yellow, the target domain is in blue, and the domain-shared module and causal structure learning module is in green.}
\label{fig2}
\end{figure} 
Fig.~\ref{fig2} shows the architecture of CDCOR, it is an end-to-end feed-forward neural network that contains three components, with inputs of observed user-item interaction records and outputs of interaction predictions. 
These three kinds of components include domain-specific module, domain-shared module and causal structure learning module.

The domain-specific module consists of two parts: the source domain part and the target domain part, they are responsible for embedding user ID and item ID into real-valued vectors of domain-specific latent features, and make the final interaction prediction. The goal of the domain-shared module is to extract users' domain-shared preferences through domain adversarial training. The causal structure learning module is designed to model the causal structure in order to get the causal invariant preferences of the user.
\subsection{Domain-Specific Embedding}
Since the notations in the source domain are similar to the target domain, for the sake of descriptive brevity, we use the target domain as an illustrative example in the subsequent description.
Firstly, we map user IDs and item IDs into a high-dimensional space to facilitate learning causal structure and similarity calculation. 
In both domains, each interaction record contain a user ID and a item ID, we use one-hot vectors to encode the sparse user and item representations as $\bm{o}_u\in\{0,1\}^{m}$ and $\bm{o}^t_i\in\{0,1\}^{n^t}$. 
Specifically, for items, we map their one-hot vectors into latent embeddings $\bm{i}^t\in\mathbb{R}^k$ with a item embedding encoder:
\begin{equation}
\bm{i}^t=E^t_i(\bm{o}^t_i;\theta^t_i)=\bm{W}^t_i\bm{o}^t_i\in\mathbb{R}^k\label{item}
\end{equation}
where $\bm{W}^t_i\in\mathbb{R}^{k\times{n^t}}$ is a learnable mapping matrix.
And for users, as mentioned earlier, we encode the user's one-hot vectors $\bm{o}_u\in\{0,1\}^{m}$ into latent attribute vectors for generate domain-specific user preference and causal learning:
\begin{equation}
\bm{u}^t_{att}=E^t_{att}(\bm{o}_u;\theta^t_{att})=\bm{W}^t_{att}\bm{o}_u\in\mathbb{R}^k\label{latent attribution}
\end{equation}
Similar to item, we use user domain-specific preference encoder to convert users' latent attribute vectors into user domain-specific preference for prediction:
\begin{equation}
\bm{u}^t=E^t_u(\bm{u}^t_{att};\theta^t_u)=\bm{W}^t_u\bm{u}^t_{att}\in\mathbb{R}^k\label{user}
\end{equation}
where $\bm{W}^t_{att}\in\mathbb{R}^{k\times{m}}$ and $\bm{W}^t_u\in\mathbb{R}^{k\times{k}}$ are learnable mapping matrices.
\subsection{Domain-Shared Embedding}
Learning user domain-shared preferences presents a challenge because user's common preferences are similar but not identical across domains.
For example, a user may care more about the plot of a novel, but care more about special effects of a movie.
To address this problem, we employ domain adversarial training method~\cite{ganin2016domain}, which is widely used to align different domains, for the purpose of extracting users' domain-shared preferences.
This part is trained by data from target and source domains.
Rich source domain data can help model training to get better parameters, and model performance in the target domain benefits as a result.
We feed user latent attribute vectors $\bm{u}_{att}$ from both domains into the encoder $E^c$ to generate user domain-shared preference embeddings:
\begin{equation}
\bm{u}^c=E^c(\bm{u}_{att};\theta^c)=g(\bm{W}^c_u\bm{u}_{att})\in\mathbb{R}^k\label{user shared}
\end{equation}
where $\bm{W}^c_u\in\mathbb{R}^{k\times{k}}$, $g$ is the activation function set as \emph{ReLU}.

To ensure $\bm{u}^c$ to be domain-shared, we also employ a domain discriminator $D$ to identify which domain the user domain-shared preference embeddings are coming from:
\begin{equation}
\hat{l}=D(\bm{u}^c;\theta^d)=\sigma(\bm{W}^d\bm{u}^c)\in\mathbb{R}^2\label{discriminator}
\end{equation}
where $\hat{l}$ is the domain label, which is a 2-dimensional vector, where the dimension 0 stands for the generated probabilities for the source domain and 1 stands for the target domain.
$\sigma$ presents the \emph{Sigmoid} function.
The loss of domain-shared part is defined as:
\begin{equation}
L_c(\theta^d,\theta^c)=-\sum^N_{n=1}l_n\log(\hat{l}_n)+(1-l_n)\log(1-\hat{l}_n)\label{loss dis}
\end{equation}

If the discriminator can't distinguish which domain the embeddings are coming from, then the embeddings can be viewed as domain-shared. 
In order to obtain domain-shared embeddings, the encoder and the discriminator play a min-max game.
On the one hand, the discriminator needs to minimise the classification loss to ensure that it can distinguish the source of the embeddings, and on the other hand, the encoder needs to maximise the loss to confuse the discriminator.
This strategic interplay results in the extraction of domain-shared preferences.
\subsection{Causal Structure Learning}
In this section, we will describe how we use DAG to learn causal structure between user attributes and user preferences.

Causal structure learning is the problem of learning graph $\mathcal{G}$ using user-item interaction records. 
In this work, we encode a graph $\mathcal{G}$ with $2k$ nodes as a weighted adjacency matrix $\bm{A}\in\mathbb{R}^{2k\times{2k}}$, $\bm{A}_{ij}$ representing the causal effect of node $i$ on node $j$.
Each node represents one dimension in the embedding of user latent attributes or user preferences.
We can formulate the Structural Causal Model (SCM) as:
\begin{equation}
\bm{H}=\bm{A}^T\bm{H}+\bm{\epsilon},\quad\bm{\epsilon}\sim\mathcal{N}(0,\bm{I})\label{scm}
\end{equation}
where $\bm{H}=\bm{u}_{att}||\bm{u}^c\in\mathbb{R}^{2k}$, $||$ is concatenate operation.
Equation (\ref{scm}) describes how children nodes are generated by their parental nodes.
When Equation (\ref{scm}) holds, the graph $\mathcal{G}$ correctly expresses the causal relationship between user latent attributes and user preferences.
There are two benefits of using latent attributes for causal learning: 1) the model does not rely on the explicit attributes of the user, so there are no privacy issues that would result in limiting the use of the model; 2) the model can mine potential user attributes from the data, allowing us to model causal structures more accurately.
We can achieve causal structure learning by minimise the following loss function:
\begin{equation}
L_{rec}=\frac{1}{N}\sum^N_{i=1}||\bm{H}_i-\bm{A}^T\bm{H}_i||^2_2\label{rec loss}
\end{equation}

We know that the causal graph can be formulated as a Bayesian network, so we need a constraint~\cite{zheng2018dags} to ensure that the adjacency matrix $\bm{A}$ is acyclic:
\begin{equation}
\mathrm{Tr}(e^{\bm{A}\odot \bm{A}})-k=0\label{dag}
\end{equation}
where $\odot$ denotes the elementwise product.
In summary, the loss of the causal part as
\begin{equation}
\begin{aligned}
\tilde{L}_{cau}(\theta^{cau})=&L_{rec}+\gamma_1 L_{dag}+\gamma_2 ||\bm{A}||_1\\
=&\frac{1}{N}\sum^N_{i=1}||\bm{H}_i-\bm{A}^T\bm{H}_i||^2_2+\gamma_1(\mathrm{Tr}(e^{\bm{A}\odot \bm{A}})-k)+\gamma_2 ||\bm{A}||_1\label{cau loss}
\end{aligned}
\end{equation}

Due to the special property of recommender systems, we add two additional constraints similar to~\cite{he2022causpref}:
\begin{itemize}
\item All paths can only from user attribute nodes to user preference nodes.
\item User preference nodes can't be the root nodes.
\end{itemize}

The above two constraints are consistent with the intuition that user preferences are generated by user attributes.
In the end, we get the total loss of the causal part as:
\begin{equation}
\begin{aligned}
L_{cau}(\theta^{cau})=&L_{rec}+\gamma_1 L_{dag}+\gamma_2L_{a2p}+\gamma_3L_{pnr}+\gamma_4 ||\bm{A}||_1\\
=&\frac{1}{N}\sum^N_{i=1}||\bm{H}_i-\bm{A}^T\bm{H}_i||^2_2+\gamma_1(\mathrm{Tr}(e^{\bm{A}\odot \bm{A}})-k)\\
&+\gamma_2 ||\bm{A}_{[k+1:2k,1:k]}||_1+\gamma_3\sum^{2k}_{i=k+1}-\log||A_{[:,i]}||_1+\gamma_4 ||\bm{A}||_1\label{cau total loss}
\end{aligned}
\end{equation}
where $L_{a2p}$ forces edges in $\bm{A}$ only from user attribute nodes to user preference nodes, and $L_{pnr}$ ensure that user preference nodes are not the root nodes.

Note that after training, we only need to input $\bm{u}_{att}$, while $\bm{u}^c$ is replaced by the zero vector $\bm{0}\in\mathbb{R}^k$, and in the output $\bm{\hat{H}}$ of the causal module we intercept the posterior $k$-dimensional vector as the final causal invariant preference $\bm{u}^{cau}$.
Since we modelled the process of generating user preferences with $\bm{A}$, we can directly infer user preferences from user attributes for correct out-of-distribution recommendations.
\subsection{Prediction and Training}
After the above steps, we design a predictor $f$ for each domain to predict the interaction probability.
We first obtain the user domain-specific preference embedding $\bm{u}^t$ and domain-shared causal invariant preference embedding $\bm{u}^{cau}$.
Then, we concatenate them and feed it into a fusion layer $h^t$ to get the final user preference embedding $\bm{u}^{ct}$.
The fusion layer acts as a attention layer, which can retain both the user domain-specific preferences and domain-shared preferences, and balance the importance between them.
Finally, we feed it with the item embeddings $\bm{i}^t$ into predictor, the interaction probability are calculated as follows:
\begin{equation}
\hat{y}^t_{ui}=f^t(h^t(\bm{u}^t||\bm{u}^{cau})\odot\bm{i}^t)=\bm{W}^t_f(\bm{u}^{ct}\odot\bm{i}^t)\in\mathbb{R}^2\label{predict}
\end{equation}
where $\bm{W}^t_f\in\mathbb{R}^{2\times{k}}$.

Considering the binary property of implicit feedback, we use cross-entropy loss to train this model:
\begin{equation}
L_t(\theta^t)=-\sum y^t_{ui}\log(\hat{y}^t_{ui})+(1-y^t_{ui})\log(1-\hat{y}^t_{ui})
\end{equation}
Then we get the overall loss function:
\begin{equation}
\begin{aligned}
L(\theta^t,\theta^s,\theta^d,\theta^c,\theta^{cau})=&L_t(\theta^t)+\lambda_1L_s(\theta^s)+\lambda_2L_c(\theta^d,\theta^c)\\
&+\lambda_3L_{cau}(\theta^{cau})+\lambda_4||\theta||_2
\end{aligned}
\end{equation}
where $\theta$ represents all trainable parameters and $\lambda$ are used to adjust the importance of each term.
Then the optimization objective can be expressed as:
\begin{equation}
\begin{aligned}
(\hat{\theta}^t,\hat{\theta}^s,\hat{\theta}^d,\hat{\theta}^{cau})&=\mathop{\arg\min}\limits_{\theta^t,\theta^s,\theta^d,\theta^{cau}}L(\theta^t,\theta^s,\theta^d,\theta^c,\theta^{cau})\\
\hat{\theta}^c&=\mathop{\arg\max}\limits_{\theta^c}L(\theta^t,\theta^s,\theta^d,\theta^c,\theta^{cau})
\end{aligned}
\end{equation}

We use gradient reversal layer (GRL)~\cite{ganin2016domain} to optimise our loss function. 
It plays different roles in forward and backward propagation. 
It acts as an identity function in forward propagation and inverts the gradient in backward propagation, so it can be optimise the parameters to make the discriminator more accurate, while the gradient reversal leads to optimisation of the preference encoder in the opposite direction, resulting in the discriminator having difficulty in distinguishing which domain the preferences are come from.
The GRL layer can be described as:
\begin{equation}
\begin{aligned}
R(x)&=x\quad\text{(forward)}\\
dR(x)/dx&=-\lambda\bm{I}\quad\text{(backward)}\label{GRL}
\end{aligned}
\end{equation}
where $\bm{I}$ is identity martix.
The GRL is placed before the discriminator after the preference encoder.
Therefore, we get a new optimization objective as:
\begin{equation}
\begin{aligned}
(\hat{\theta}^t,\hat{\theta}^s,\hat{\theta}^d,\hat{\theta}^c,\hat{\theta}^{cau})=\mathop{\arg\min}\limits_{\theta^t,\theta^s,\theta^d,\theta^c,\theta^{cau}}L(\theta^t,\theta^s,\theta^d,\theta^c,\theta^{cau})
\end{aligned}
\end{equation}

Consequently, the model parameters can be optimized by SGD-like algorithms.
\section{Experiments}
This section presents the results of the experiment on two real-world datasets, and we aim to answer the following questions:
\begin{itemize}
\item \textbf{RQ1:} How does the proposed model CDCOR perform compared with state-of-the-art baselines? 
\item \textbf{RQ2:} How does the cross-domain module and causal learning module affect the performance?
\item \textbf{RQ3:} How robust is CDCOR to varying degrees of distribution shift?
\item \textbf{RQ4:} Can our proposed model solve the data sparsity problem in OOD recommendation?
\end{itemize}

\subsection{Experimental Setup}
\subsubsection{Datasets.}We use two widely used cross-domain recommendation datasets Douban\nocite{zhu2020graphical}\footnote{\url{https://github.com/FengZhu-Joey/GA-DTCDR}} and Tenrec\nocite{yuan2022tenrec}\footnote{\url{https://github.com/yuangh-x/2022-NIPS-Tenrec}}.
Table~\ref{tab1} shows the statistics of the two datasets.
For Douban dataset, we use the movie domain, which has rich interaction records as the source domain and the book domain is the target domain, which has few interaction records. 
\begin{table}
\renewcommand\arraystretch{1.1}
\caption{The statistics of the datasets}
\begin{center}
\begin{tabular}{ccccc}
\hline
Dataset&Domain&\#Users&\#Items&\#Interactions\\
\hline
\multirow{2}*{Douban}&movie(s)&\multirow{2}*{2106}&9555&969937\\
~&book(t)&~&6777&95974\\
\multirow{2}*{Tenrec}&video(s)&\multirow{2}*{2151}&37041&247942\\
~&article(t)&~&2869&37641\\
\hline
\end{tabular}
\label{tab1}
\end{center}
\end{table}
This dataset contains user IDs, item IDs and ratings.
And for Tenrec, the source domain is the video domain and the article domain is used as the target domain. It contains not only the user IDs, item IDs and ratings, but also the user's gender.
Note that the ratings are explicit feedback, we transformed it into implicit feedback, where the interactions with ratings $\geq$ 4 as positive samples. 

On both datasets we first perform experiments with independent and identically distributed (IID) scenarios, where both the training and testing data are obtained by random sampling.
Then we set up two OOD settings as flollow:
1) \textbf{User Degree Bias}: Active users in the platform interact with a wide range of different types of item, the active users will have smoother feature embeddings, which will lead to incorrect recommendation result.
In this setting, we randomly sample training data from raw dataset, and sample users mainly with the upper degree for testing data.
2) \textbf{User Attribute Bias}: Attributes of users change over time, for example, the user's income will increase.
In this setting, we divide the dataset into two parts according to user's gender, then sample from raw dataset for training, and adjust the different ratios of data from two parts for test.

The user degree bias setting (OOD\#1) is experimented on both datasets, and the user attribute bias setting (OOD\#2) is only performed on the Tenrec dataset.
In all settings, the proportion of training, validation and testing data are set as 8:1:1.

\subsubsection{Baselines.}We compare our CDCOR with eight state-of-the-art recommendation methods: 1) Traditional recommendations: GMF~\cite{he2017neural}, IPS~\cite{schnabel2016recommendations}. 2) Cross-domain recommendations: DARec~\cite{yuan2019darec}, HMRec~\cite{liao2022heterogeneous}, PTUPCDR~\cite{zhu2022personalized}. 3) Disentangling based OOD recommendations: DICE~\cite{zheng2021disentangling}, DCCL~\cite{zhao2023disentangled}. 4) Causal based OOD recommendation: CausPref~\cite{he2022causpref}.

\subsubsection{Metrics and Implementation Details.}
In this paper, we adopt Hit Ratio (HR) and Normalized Discounted Cumulative Gain (NDCG) as the metrics, which are widely used in recommendation senarios.
The HR@\textit{k} is the ratio of the test item appears in the top-\textit{k} list, while the NDCG@\textit{k} consider the precision position of the hit, with higher scores for higher positions.
We use the \textit{leave-one-out} evaluate strategy~\cite{rendle2012bpr,he2016fast}, which sample 99 negative items for each positive item in the test set.

We set the embedding dimensionality as 16, learning rate as 0.01 for all scenarios.
We design two hidden layers for domain discriminator, and one hidden layer for other MLP.
Then we set the trade-off parameters $\lambda_1$, $\lambda_2$, $\lambda_3$ and $\lambda_4$ as 1, 0.5, 1, 0.00001, respectively.
Hyperparameters of all baselines are set to optimal values.
Finally, we report the average result of five random times.
\subsection{Performance Comparison (RQ1 \& RQ2)}
Table~\ref{tab2}, ~\ref{tab3} and ~\ref{tab4} show the results on two datasets including two IID tests and three OOD tests results.
The numbers in parentheses represent the percentage of performance degradation of the model in the OOD tests.
\begin{table}
\renewcommand\arraystretch{0.8}
\centering
\caption{Performance comparison on Douban dataset}
\resizebox{\textwidth}{!}{
\begin{tabular}{c|cccc|cccc}
\toprule
\textbf{IID/OOD}&\multicolumn{4}{c|}{\textbf{IID}}&\multicolumn{4}{c}{\textbf{OOD\#1}}\\
\midrule
\textbf{Metric}&\textbf{H@5}&\textbf{H@10}&\textbf{N@5}&\textbf{N@10}&\textbf{H@5}&\textbf{H@10}&\textbf{N@5}&\textbf{N@10}\\
\midrule
GMF&0.2142&0.3209&0.1448&0.1791&0.1750(-18.29\%)&0.2793(-12.99\%)&0.1156(-20.18\%)&0.1491(-16.77\%)\\
&($\pm$0.0040)&($\pm$0.0070)&($\pm$0.0036)&($\pm$0.0043)&($\pm$0.0079)&($\pm$0.0064)&($\pm$0.0051)&($\pm$0.0038)\\
IPS&0.2139&0.3197&0.1452&0.1792&0.1833(-14.28\%)&0.2898(-9.36\%)&0.1190(-18.08\%)&0.1530(-14.62\%)\\
&($\pm$0.0032)&($\pm$0.0041)&($\pm$0.0027)&($\pm$0.0030)&($\pm$0.0060)&($\pm$0.0043)&($\pm$0.0055)&($\pm$0.0045)\\
DICE&0.2467&0.3667&0.1685&0.2073&0.2089(-15.31\%)&0.3127(-14.71\%)&0.1391(-17.49\%)&0.1724(-16.82\%)\\
&($\pm$0.0090)&($\pm$0.0054)&($\pm$0.0060)&($\pm$0.0042)&($\pm$0.0059)&($\pm$0.0037)&($\pm$0.0045)&($\pm$0.0042)\\
DCCL&\textbf{0.2641}&\textbf{0.3850}&\textbf{0.1797}&\textbf{0.2186}&\underline{0.2265(-14.27\%)}&\underline{0.3516(-8.68\%)}&\underline{0.1461(-18.73\%)}&\underline{0.1863(-14.76\%)}\\
&($\pm$0.0160)&($\pm$0.0145)&($\pm$0.0137)&($\pm$0.0131)&($\pm$0.0047)&($\pm$0.0044)&($\pm$0.0024)&($\pm$0.0021)\\
DARec&0.2373&0.3437&0.1646&0.1988&0.1968(-17.05\%)&0.3014(-12.31\%)&0.1329(-19.24\%)&0.1665(-16.23\%)\\
&($\pm$0.0039)&($\pm$0.0039)&($\pm$0.0031)&($\pm$0.0027)&($\pm$0.0067)&($\pm$0.0028)&($\pm$0.0046)&($\pm$0.0026)\\
HMRec&0.2408&0.3418&0.1666&0.1991&0.2038(-15.35\%)&0.3060(-10.46\%)&0.1405(-15.66\%)&0.1733(-12.95\%)\\
&($\pm$0.0073)&($\pm$0.0057)&($\pm$0.0052)&($\pm$0.0046)&($\pm$0.0113)&($\pm$0.0090)&($\pm$0.0080)&($\pm$0.0073)\\
PTUPCDR&0.2355&0.3444&0.1612&0.1963&0.2048(-13.04\%)&0.3096(-10.10\%)&0.1370(-15.01\%)&0.1708(-13.00\%)\\
&($\pm$0.0038)&($\pm$0.0029)&($\pm$0.0022)&($\pm$0.0015)&($\pm$0.0070)&($\pm$0.0088)&($\pm$0.0050)&($\pm$0.0052)\\
CausPref&0.2237&0.3282&0.1542&0.1878&0.1853(-17.18\%)&0.2894(-11.80\%)&0.1255(-18.62\%)&0.1589(-15.41\%)\\
&($\pm$0.0032)&($\pm$0.0010)&($\pm$0.0033)&($\pm$0.0025)&($\pm$0.0058)&($\pm$0.0052)&($\pm$0.0028)&($\pm$0.0020)\\
\midrule
CDCOR&\underline{0.2581}&\underline{0.3782}&\underline{0.1779}&\underline{0.2166}&\textbf{0.2362(-8.51\%)}&\textbf{0.3574(-5.50\%)}&\textbf{0.1582(-11.05\%)}&\textbf{0.1973(-8.90\%)}\\
&($\pm$0.0035)&($\pm$0.0027)&($\pm$0.0036)&($\pm$0.0030)&($\pm$0.0074)&($\pm$0.0047)&($\pm$0.0056)&($\pm$0.0043)\\
w/o causal&0.2560&0.3748&0.1750&0.2132&0.2215(-13.46\%)&0.3388(-9.61\%)&0.1467(-16.15\%)&0.1844(-13.53\%)\\
&($\pm$0.0046)&($\pm$0.0053)&($\pm$0.0046)&($\pm$0.0049)&($\pm$0.0070)&($\pm$0.0046)&($\pm$0.0067)&($\pm$0.0056)\\
w/o source&0.2542&0.3697&0.1744&0.2116&0.2246(-11.65\%)&0.3375(-8.71\%)&0.1493(-14.35\%)&0.1860(-12.09\%)\\
&($\pm$0.0049)&($\pm$0.0043)&($\pm$0.0025)&($\pm$0.0025)&($\pm$0.0093)&($\pm$0.0077)&($\pm$0.0071)&($\pm$0.0063)\\
\bottomrule
\end{tabular}}
\label{tab2}
\end{table}
\begin{table}\tiny
\centering
\caption{Performance comparison on Tenrec dataset (IID)}
\begin{tabular}{c|cccc}
\toprule
\textbf{IID/OOD}&\multicolumn{4}{c}{\textbf{IID}}\\
\midrule
\textbf{Metric}&\textbf{H@5}&\textbf{H@10}&\textbf{N@5}&\textbf{N@10}\\
\midrule
GMF&0.4422&0.6238&0.3077&0.3664\\
&($\pm$0.0089)&($\pm$0.0068)&($\pm$0.0074)&($\pm$0.0063)\\
IPS&0.4425&0.6292&0.3074&0.3678\\
&($\pm$0.0044)&($\pm$0.0058)&($\pm$0.0028)&($\pm$0.0019)\\
DICE&0.4698&\textbf{0.6993}&0.3299&\underline{0.4043}\\
&($\pm$0.0098)&($\pm$0.0102)&($\pm$0.0089)&($\pm$0.0088)\\
DCCL&\underline{0.4926}&0.6536&\textbf{0.3475}&0.3997\\
&($\pm$0.0041)&($\pm$0.0036)&($\pm$0.0027)&($\pm$0.0028)\\
DARec&0.4746&0.6615&0.3302&0.3905\\
&($\pm$0.0044)&($\pm$0.0091)&($\pm$0.0034)&($\pm$0.0054)\\
HMRec&0.4610&0.6483&0.3182&0.3787\\
&($\pm$0.0105)&($\pm$0.0096)&($\pm$0.0116)&($\pm$0.0101)\\
PTUPCDR&0.4670&0.6531&0.3199&0.3800\\
&($\pm$0.0059)&($\pm$0.0030)&($\pm$0.0065)&($\pm$0.0047)\\
CausPref&0.4386&0.6260&0.3012&0.3618\\
&($\pm$0.0078)&($\pm$0.0024)&($\pm$0.0052)&($\pm$0.0034)\\
\midrule
CDCOR&\textbf{0.4967}&\underline{0.6878}&\underline{0.3452}&\textbf{0.4070}\\
&($\pm$0.0083)&($\pm$0.0065)&($\pm$0.0044)&($\pm$0.0031)\\
w/o causal&0.4717&0.6668&0.3243&0.3873\\
&($\pm$0.0097)&($\pm$0.0102)&($\pm$0.0080)&($\pm$0.0079)\\
w/o source&0.4867&0.6718&0.3385&0.3983\\
&($\pm$0.0069)&($\pm$0.0054)&($\pm$0.0068)&($\pm$0.0050)\\
\bottomrule
\end{tabular}
\label{tab3}
\end{table}
\begin{table}
\centering
\caption{Performance comparison on Tenrec dataset (OOD)}
\resizebox{\textwidth}{!}{
\begin{tabular}{c|cccc|cccc}
\toprule
\textbf{IID/OOD}&\multicolumn{4}{c|}{\textbf{OOD\#1}}&\multicolumn{4}{c}{\textbf{OOD\#2}}\\
\midrule
\textbf{Metric}&\textbf{H@5}&\textbf{H@10}&\textbf{N@5}&\textbf{N@10}&\textbf{H@5}&\textbf{H@10}&\textbf{N@5}&\textbf{N@10}\\
\midrule
GMF&0.3671(-16.99\%)&0.5444(-12.72\%)&0.2446(-20.51\%)&0.3017(-17.64\%)&0.4050(-8.43\%)&0.5888(-5.61\%)&0.2828(-8.10\%)&0.3421(-6.63\%)\\
&($\pm$0.0101)&($\pm$0.0136)&($\pm$0.0075)&($\pm$0.0074)&($\pm$0.0144)&($\pm$0.0060)&($\pm$0.0123)&($\pm$0.0092)\\
IPS&0.3721(-15.91\%)&0.5453(-13.34\%)&0.2497(-18.76\%)&0.3057(-16.88\%)&0.4118(-6.94\%)&0.5816(-7.56\%)&0.2859(-7.00\%)&0.3406(-7.41\%)\\
&($\pm$0.0083)&($\pm$0.0031)&($\pm$0.0065)&($\pm$0.0044)&($\pm$0.0209)&($\pm$0.0059)&($\pm$0.0142)&($\pm$0.0093)\\
DICE&0.4060(-13.58\%)&\underline{0.6066(-13.26\%)}&0.2736(-17.05\%)&\underline{0.3389(-16.18\%)}&0.4514(-3.39\%)&\underline{0.6601(-5.60\%)}&0.3180(-3.60\%)&0.3856(-4.63\%)\\
&($\pm$0.0144)&($\pm$0.0126)&($\pm$0.0128)&($\pm$0.0121)&($\pm$0.0193)&($\pm$0.0082)&($\pm$0.0165)&($\pm$0.0122)\\
DCCL&\underline{0.4196(-14.81\%)}&0.5927(-9.32\%)&\underline{0.2828(-18.62\%)}&0.3382(-15.41\%)&\underline{0.4792(-2.71\%)}&0.6452(-1.28\%)&\textbf{0.3365(-3.16\%)}&\underline{0.3902(-2.38\%)}\\
&($\pm$0.0113)&($\pm$0.0082)&($\pm$0.0095)&($\pm$0.0057)&($\pm$0.0099)&($\pm$0.0160)&($\pm$0.0059)&($\pm$0.0073)\\
DARec&0.3886(-18.11\%)&0.5644(-14.68\%)&0.2643(-19.95\%)&0.3207(-17.89\%)&0.4263(-10.19\%)&0.6289(-4.93\%)&0.2957(-10.43\%)&0.3607(-7.64\%)\\
&($\pm$0.0165)&($\pm$0.0133)&($\pm$0.0128)&($\pm$0.0119)&($\pm$0.0100)&($\pm$0.0032)&($\pm$0.0066)&($\pm$0.0043)\\
HMRec&0.3950(-14.32\%)&0.5781(-10.83\%)&0.2661(-16.35\%)&0.3252(-14.13\%)&0.4291(-6.92\%)&0.6143(-5.24\%)&0.2918(-8.29\%)&0.3516(-7.15\%)\\
&($\pm$0.0197)&($\pm$0.0110)&($\pm$0.0152)&($\pm$0.0115)&($\pm$0.0202)&($\pm$0.0181)&($\pm$0.0153)&($\pm$0.0148)\\
PTUPCDR&0.3860(-17.35\%)&0.5771(-11.63\%)&0.2628(-17.87\%)&0.3226(-15.11\%)&0.4131(-11.55\%)&0.5998(-8.16\%)&0.2884(-9.87\%)&0.3487(-8.24\%)\\
&($\pm$0.0153)&($\pm$0.0115)&($\pm$0.0046)&($\pm$0.0054)&($\pm$0.0031)&($\pm$0.0077)&($\pm$0.0071)&($\pm$0.0064)\\
CausPref&0.3256(-25.75\%)&0.4984(-20.39\%)&0.2179(-27.65\%)&0.2736(-24.39\%)&0.3836(-12.54\%)&0.5600(-10.55\%)&0.2639(-12.40\%)&0.3206(-11.40\%)\\
&($\pm$0.0319)&($\pm$0.0446)&($\pm$0.0283)&($\pm$0.0300)&($\pm$0.0113)&($\pm$0.0093)&($\pm$0.0062)&($\pm$0.0041)\\
\midrule
CDCOR&\textbf{0.4357(-12.27\%)}&\textbf{0.6310(-8.25\%)}&\textbf{0.2954(-14.44\%)}&\textbf{0.3588(-11.84\%)}&\textbf{0.4880(-1.74\%)}&\textbf{0.6841(-0.54\%)}&\underline{0.3363(-2.57\%)}&\textbf{0.3997(-1.77\%)}\\
&($\pm$0.0111)&($\pm$0.0117)&($\pm$0.0085)&($\pm$0.0088)&($\pm$0.0073)&($\pm$0.0060)&($\pm$0.0047)&($\pm$0.0042)\\
w/o causal&0.3991(-15.39\%)&0.5793(-13.13\%)&0.2619(-19.24\%)&0.3206(-17.23\%)&0.4561(-3.29\%)&0.6476(-2.89\%)&0.3089(-4.73\%)&0.3704(-4.36\%)\\
&($\pm$0.0137)&($\pm$0.0138)&($\pm$0.0058)&($\pm$0.0052)&($\pm$0.0194)&($\pm$0.0067)&($\pm$0.0134)&($\pm$0.0088)\\
w/o source&0.4111(-15.52\%)&0.6026(-10.30\%)&0.2764(-18.33\%)&0.3380(-15.13\%)&0.4757(-2.25\%)&0.6637(-1.20\%)&0.3252(-3.92\%)&0.3858(-3.12\%)\\
&($\pm$0.0096)&($\pm$0.0079)&($\pm$0.0089)&($\pm$0.0082)&($\pm$0.0127)&($\pm$0.0063)&($\pm$0.0082)&($\pm$0.0037)\\
\bottomrule
\end{tabular}}
\label{tab4}
\end{table}
We have the following observations:
1) The performance of all the models have a significant degradation in the OOD test, which is due to the shift of the data distribution under the OOD environment. 
2) CDCOR has a comparable performance with the SOTA method in the IID test, and has the best performance in the OOD test. It is worth noting that CDCOR has the least performance degradation under the OOD test, it verifies that CDCOR has a good generalisation performance when facing the distribution shift, which indicates that correctly capturing causal invariance in the data distribution is beneficial for improving the performance of OOD recommendations in recommender systems. 
3) Compared with CausPref, CDCOR has a huge advantage in both IID and OOD tests, suggesting that information from other domains not only improves the model recommendation performance, but also helps the model to learn the correct causal structure. 
4) IPS has been shown to be useful in some of these cases, while not in others due to the fact that IPS needs to rely on unbiased data that is not always available.
5) Three cross-domain based recommendation methods PTUPCDR, HMRec and DARec perform close to each other and outperform traditional single-domain recommendation methods, but still underperform in OOD tests, suggesting that relying only on cross-domain data is difficult to improve model performance in OOD environments.
6) DICE and DCCL two methods based on the disentangling of embedding have good performance in both IID and OOD tests, but the performance drops more in OOD test than CDCOR, indicating that the decoupling of the factors for user interactions is helpful for recommendation, but can only be used for a specific distribution shift, which may require more fine-grained disentangling when faces with more complex distribution shift environments.
7) The two ablation variants w/o causal (removed the causal inference part) and w/o source (removed the source part) do not perform as well as CDCOR in both the IID and OOD tests. Their performance drop is more noticeable in the OOD test, which shows that correctly capturing causal invariance and the knowledge from other domains both play an important role in improving the out-of-distribution generalisation performance of recommender systems.

\subsection{Robustness experiments (RQ3)}
We test the stability of each model under varying levels of distribution shift.
In user degree bias setting, we fix the ratio of the two types user in the training dataset to be 4:6 and adjusted it to 7:3 in the testing dataset.
In user attribute bias setting, this ratio is 8:2 and 2:8 in the training and test sets, respectively.
\begin{figure}
\centering
\subfloat[]{
    \includegraphics[width=0.325\linewidth]{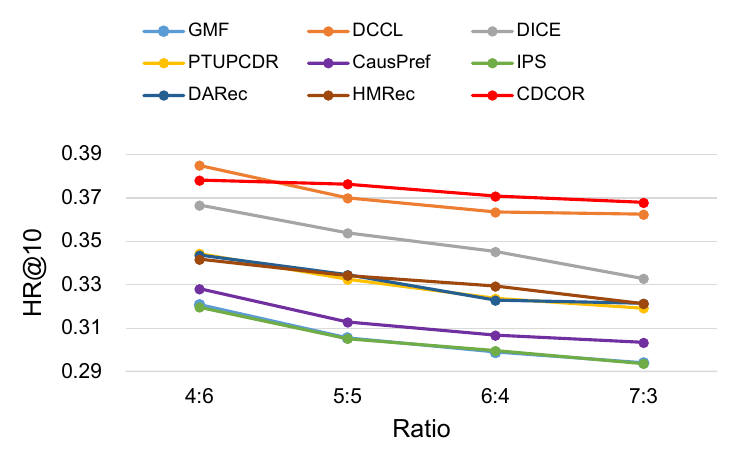}
}
\subfloat[]{
    \includegraphics[width=0.325\linewidth]{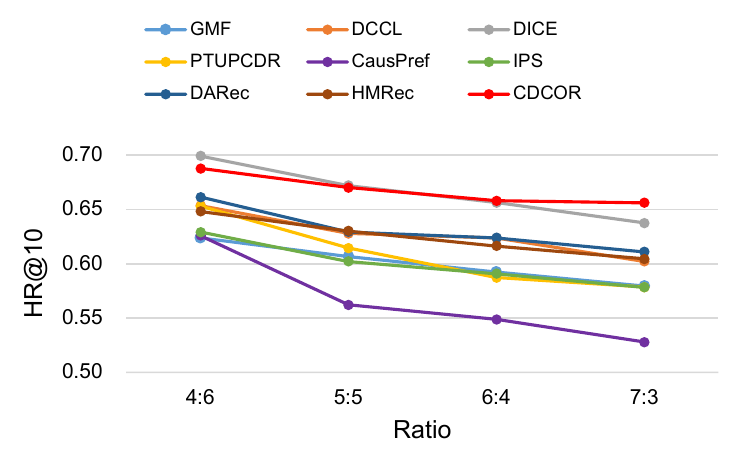}
}
\subfloat[]{
    \includegraphics[width=0.325\linewidth]{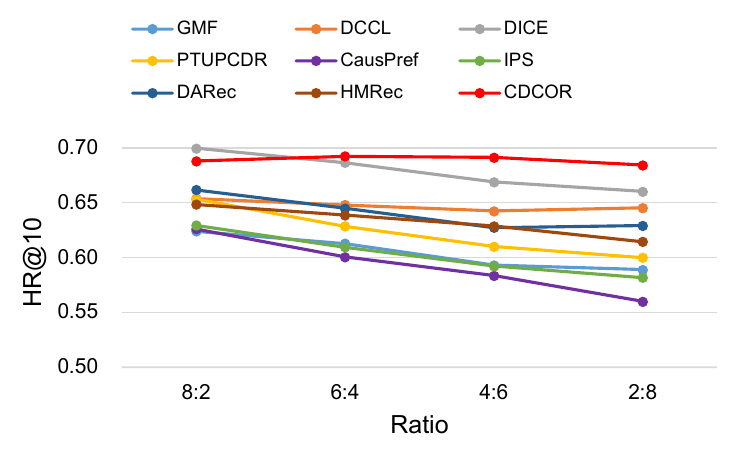}
}
\caption{Robustness experiments. (a) OOD\#1 test in Douban. (b) OOD\#1 test in Tenrec. (c) OOD\#2 test in Tenrec.}
\label{robust}
\end{figure}

Fig.~\ref{robust} reports the experiment results. 
We observe that all methods show a decrease when facing the distribution shift.
In the IID case, CDCOR is comparable to the best performing method.
The advantage of CDCOR becomes more and more obvious as the distribution shift increases, which verifies that CDCOR possesses better stability and stronger robustness in various OOD environments.

\subsection{Performance Against Sparsity (RQ4)}
\begin{figure}
\centering
\subfloat[]{
    \includegraphics[width=0.325\linewidth]{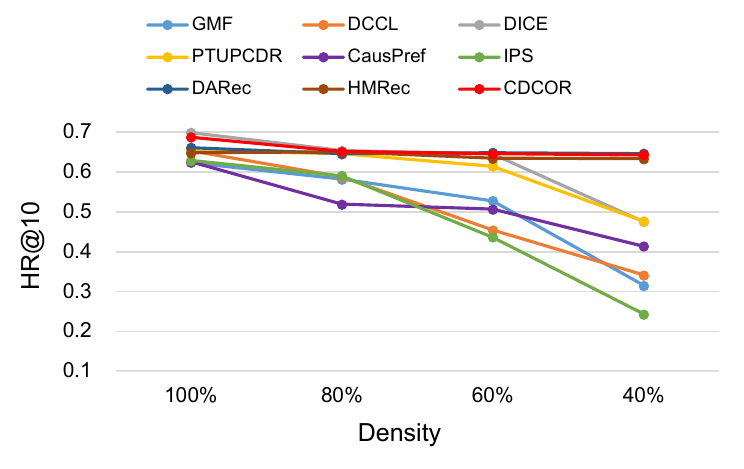}
}
\subfloat[]{
    \includegraphics[width=0.325\linewidth]{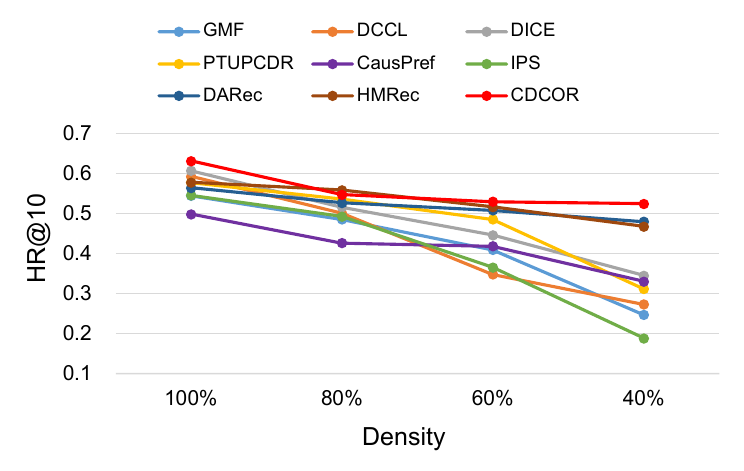}
}
\subfloat[]{
    \includegraphics[width=0.325\linewidth]{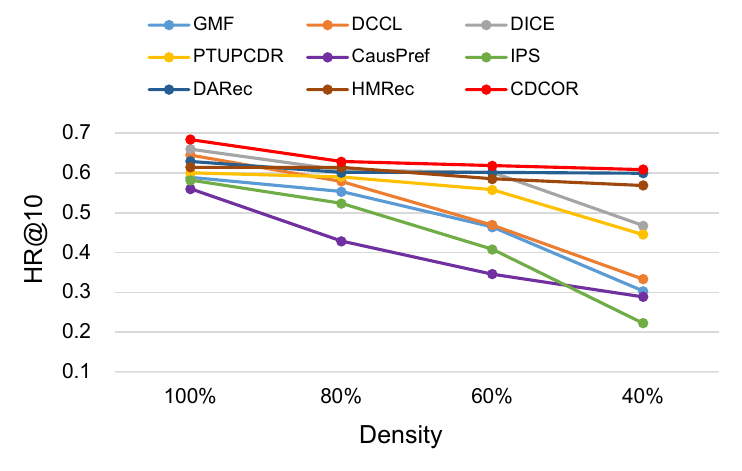}
}
\caption{Performance against sparsity in Tenrec. (a) IID test. (b) OOD\#1 test. (c) OOD\#2 test.}
\label{sparsityT}
\end{figure}

\begin{figure}
\centering
\subfloat[]{
    \includegraphics[width=0.49\linewidth]{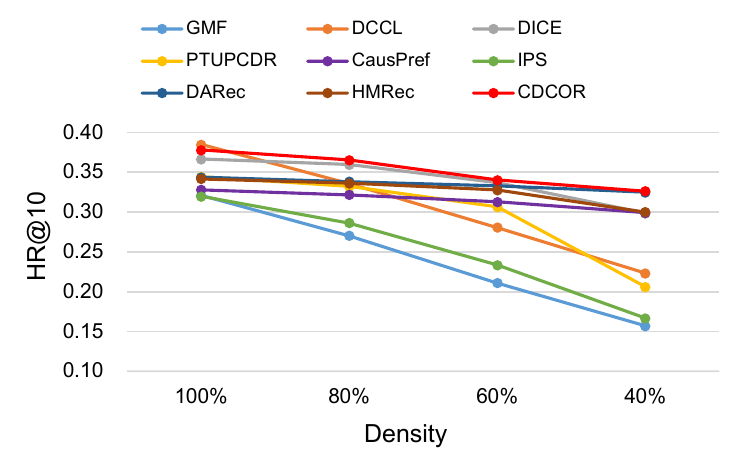}
}\hspace{-5mm}
\subfloat[]{
    \includegraphics[width=0.49\linewidth]{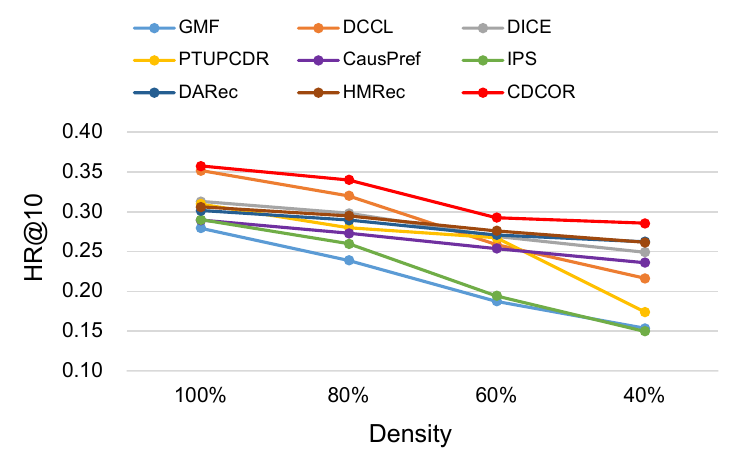}
}
\caption{Performance against sparsity in Douban. (a) IID test. (b) OOD\#1 test}
\label{sparsityD}
\end{figure}

We test the ability of each method to against sparsity in two IID settings and three OOD settings on two datasets. 
The horizontal axis indicates how much data was used for training. 
We can see from Fig.~\ref{sparsityT} and Fig.~\ref{sparsityD} that the performance of all methods decrease as the sparsity of the data increases, and decrease faster in OOD environments.
In the vast majority of cases, CDCOR outperforms the other methods, and the lower the density, the more obvious the performance advantage of CDCOR over other methods, especially in OOD tests.
The results show that CDCOR is able to learn the correct causal relationship in the case of sparse data in the target domain, thus improving OOD recommendation ability. 
This also verifies that data sparsity does affect the model to learn causal relationships.

\section{Related Work}
\subsubsection{Causal Recommendation.}
Causal inference has many applications in recommender systems, and these works are mainly used to remove the bias~\cite{wei2021model,zhang2021causal,chen2021autodebias}.
Inverse propensity weighting (IPW) is a widely used method and achieves excellent performance.
For example, RelMF~\cite{saito2020unbiased} treats the probability of an item being exposed to a user as a propensity score and proposes an unbiased estimator of the ideal loss function which is optimized by clipped propensity score to eliminate the exposure bias.

Some works focuse on beyond-accuracy objectives, such as explanability~\cite{xian2019reinforcement,tan2021counterfactual} and fairness~\cite{zhang2021counterfactual}.
For instance, Li et al.~\cite{li2021towards} causally models the recommender systems in terms of sensitive and non-sensitive features, and introduces adversarial learning to remove information about sensitive features from user embeddings to mitigate unfairness.
However, there is still limited existing works on causal inference-based recommender systems for OOD problems.

\subsubsection{Out-of-Distribution Recommendation.}
User preference may shift over time due to the changes of user attributes, which is frequent in the real-world scenarios.
Recommender systems that ignore these changes can lead to inappropriate recommendations.
Existing works that deal with user preference shifts mainly fall into two categories.
1) Decoupled recommendations enhance robustness in the distribution shift scenario by learning decoupled user preference embeddings~\cite{zheng2021disentangling,zhao2023disentangled}.
However, previous studies ignore changes in user attributes and only address specific OOD situations, can't widely use.
2) Causal inference based methods aims at modelling causal invariance~\cite{wang2022causal,he2022causpref}.
These methods can handle more variable OOD scenarios, but they require explicit attributes and dense datasets.
Our propose method can overcome these shortcomings by utilizing cross-domain knowledge.

\subsubsection{Cross-domain Recommendation.}
Cross-domain recommendation is a common approach used to address the issue of data sparsity in recommender systems~\cite{zhao2022multi,man2017cross}. 
The main idea of cross-domain recommendation is to transfer knowledge from source domain to target domain. 
Deep learning has received a lot of attention in recent years, and many researchers have adopted it into cross-domain approaches~\cite{hu2018conet,zhu2019dtcdr}.
The main idea in deep learning methods is to extract latent factors that are common to different domains through deep neural networks.
For instance, EMCDR~\cite{man2017cross} proposes the use of a multi-layer perceptron to capture cross-domain nonlinear mapping functions for user or item embeddings.

Inspired by the fact that adversarial training can be used to align two domains~\cite{tzeng2017adversarial,ganin2016domain}, many cross-domain recommender models based on adversarial training have been proposed in recent years.
DARec~\cite{yuan2019darec} draws inspiration from domain adaptation and introduces a deep domain adaptation model capable of extracting and transferring common rating patterns from rating matrices.
We extend the application of cross-domain recommendations by combining cross-domain methods with causal inference.

\section{Conclusion}
In this paper, we propose a novel model called Cross-Domain Causal Preference Learning for Out-of-Distribution Recommendation (CDCOR).
Unlike conventional causal inference-based recommendation models, our proposed model leverages knowledge from other domains to not only mitigate the effects of data sparsity to help the model learn the correct causal relationship in target domain, but also to mitigate the effects of outdated interaction records and extend causal structure learning to the latent attribute level.
At last, experiments in multiple scenarios demonstrate the superior performance of CDCOR in out-of-distribution recommendation.

\subsubsection*{Acknowledgements.}
This work is supported by National Natural Science Foundation of China under grant 61972270.

\bibliographystyle{splncs04}

\bibliography{main}

\begin{thebibliography}{10}
\providecommand{\url}[1]{\texttt{#1}}
\providecommand{\urlprefix}{URL }
\providecommand{\doi}[1]{https://doi.org/#1}

\bibitem{chen2021autodebias}
Chen, J., Dong, H., Qiu, Y., He, X., Xin, X., Chen, L., Lin, G., Yang, K.:
  Autodebias: Learning to debias for recommendation. In: SIGIR. pp. 21--30
  (2021)

\bibitem{ganin2016domain}
Ganin, Y., Ustinova, E., Ajakan, H., Germain, P., Larochelle, H., Laviolette,
  F., Marchand, M., Lempitsky, V.: Domain-adversarial training of neural
  networks. JMLR  \textbf{17}(1),  2096--2030 (2016)

\bibitem{he2017neural}
He, X., Liao, L., Zhang, H., Nie, L., Hu, X., Chua, T.S.: Neural collaborative
  filtering. In: WWW. pp. 173--182 (2017)

\bibitem{he2016fast}
He, X., Zhang, H., Kan, M.Y., Chua, T.S.: Fast matrix factorization for online
  recommendation with implicit feedback. In: SIGIR. pp. 549--558 (2016)

\bibitem{he2022causpref}
He, Y., Wang, Z., Cui, P., Zou, H., Zhang, Y., Cui, Q., Jiang, Y.: Causpref:
  Causal preference learning for out-of-distribution recommendation. In: WWW.
  pp. 410--421 (2022)

\bibitem{hu2018conet}
Hu, G., Zhang, Y., Yang, Q.: Conet: Collaborative cross networks for
  cross-domain recommendation. In: CIKM. pp. 667--676 (2018)

\bibitem{li2021towards}
Li, Y., Chen, H., Xu, S., Ge, Y., Zhang, Y.: Towards personalized fairness
  based on causal notion. In: SIGIR. pp. 1054--1063 (2021)

\bibitem{liao2022heterogeneous}
Liao, W., Zhang, Q., Yuan, B., Zhang, G., Lu, J.: Heterogeneous multidomain
  recommender system through adversarial learning. TNNLS  (2022)

\bibitem{liu2021towards}
Liu, J., Shen, Z., He, Y., Zhang, X., Xu, R., Yu, H., Cui, P.: Towards
  out-of-distribution generalization: A survey. arXiv preprint arXiv:2108.13624
   (2021)

\bibitem{man2017cross}
Man, T., Shen, H., Jin, X., Cheng, X.: Cross-domain recommendation: An
  embedding and mapping approach. In: IJCAI. vol.~17, pp. 2464--2470 (2017)

\bibitem{rendle2012bpr}
Rendle, S., Freudenthaler, C., Gantner, Z., Schmidt-Thieme, L.: Bpr: Bayesian
  personalized ranking from implicit feedback. arXiv preprint arXiv:1205.2618
  (2012)

\bibitem{saito2020unbiased}
Saito, Y., Yaginuma, S., Nishino, Y., Sakata, H., Nakata, K.: Unbiased
  recommender learning from missing-not-at-random implicit feedback. In: WSDM.
  pp. 501--509 (2020)

\bibitem{schnabel2016recommendations}
Schnabel, T., Swaminathan, A., Singh, A., Chandak, N., Joachims, T.:
  Recommendations as treatments: Debiasing learning and evaluation. In: ICML.
  pp. 1670--1679. PMLR (2016)

\bibitem{tan2021counterfactual}
Tan, J., Xu, S., Ge, Y., Li, Y., Chen, X., Zhang, Y.: Counterfactual
  explainable recommendation. In: CIKM. pp. 1784--1793 (2021)

\bibitem{tzeng2017adversarial}
Tzeng, E., Hoffman, J., Saenko, K., Darrell, T.: Adversarial discriminative
  domain adaptation. In: CVPR. pp. 7167--7176 (2017)

\bibitem{wang2022causal}
Wang, W., Lin, X., Feng, F., He, X., Lin, M., Chua, T.S.: Causal representation
  learning for out-of-distribution recommendation. In: WWW. pp. 3562--3571
  (2022)

\bibitem{wang2023causal}
Wang, W., Lin, X., Wang, L., Feng, F., Ma, Y., Chua, T.S.: Causal disentangled
  recommendation against user preference shifts. TOIS  (2023)

\bibitem{wei2021model}
Wei, T., Feng, F., Chen, J., Wu, Z., Yi, J., He, X.: Model-agnostic
  counterfactual reasoning for eliminating popularity bias in recommender
  system. In: KDD. pp. 1791--1800 (2021)

\bibitem{xian2019reinforcement}
Xian, Y., Fu, Z., Muthukrishnan, S., De~Melo, G., Zhang, Y.: Reinforcement
  knowledge graph reasoning for explainable recommendation. In: SIGIR. pp.
  285--294 (2019)

\bibitem{yuan2019darec}
Yuan, F., Yao, L., Benatallah, B.: Darec: deep domain adaptation for
  cross-domain recommendation via transferring rating patterns. In: IJCAI. pp.
  4227--4233 (2019)

\bibitem{yuan2022tenrec}
Yuan, G., Yuan, F., Li, Y., Kong, B., Li, S., Chen, L., Yang, M., Yu, C., Hu,
  B., Li, Z., et~al.: Tenrec: A large-scale multipurpose benchmark dataset for
  recommender systems. NIPS  \textbf{35},  11480--11493 (2022)

\bibitem{zhang2021counterfactual}
Zhang, X., Jia, H., Su, H., Wang, W., Xu, J., Wen, J.R.: Counterfactual reward
  modification for streaming recommendation with delayed feedback. In: SIGIR.
  pp. 41--50 (2021)

\bibitem{zhang2021causal}
Zhang, Y., Feng, F., He, X., Wei, T., Song, C., Ling, G., Zhang, Y.: Causal
  intervention for leveraging popularity bias in recommendation. In: SIGIR. pp.
  11--20 (2021)

\bibitem{zhao2023disentangled}
Zhao, W., Tang, D., Chen, X., Lv, D., Ou, D., Li, B., Jiang, P., Gai, K.:
  Disentangled causal embedding with contrastive learning for recommender
  system. In: WWW. pp. 406--410 (2023)

\bibitem{zhao2022multi}
Zhao, X., Yang, N., Yu, P.S.: Multi-sparse-domain collaborative recommendation
  via enhanced comprehensive aspect preference learning. In: WSDM. pp.
  1452--1460 (2022)

\bibitem{zheng2018dags}
Zheng, X., Aragam, B., Ravikumar, P.K., Xing, E.P.: Dags with no tears:
  Continuous optimization for structure learning. NIPS  \textbf{31} (2018)

\bibitem{zheng2021disentangling}
Zheng, Y., Gao, C., Li, X., He, X., Li, Y., Jin, D.: Disentangling user
  interest and conformity for recommendation with causal embedding. In: WWW.
  pp. 2980--2991 (2021)

\bibitem{zhu2019dtcdr}
Zhu, F., Chen, C., Wang, Y., Liu, G., Zheng, X.: Dtcdr: A framework for
  dual-target cross-domain recommendation. In: CIKM. pp. 1533--1542 (2019)

\bibitem{zhu2020graphical}
Zhu, F., Wang, Y., Chen, C., Liu, G., Zheng, X.: A graphical and attentional
  framework for dual-target cross-domain recommendation. In: IJCAI. pp.
  3001--3008 (2020)

\bibitem{zhu2022personalized}
Zhu, Y., Tang, Z., Liu, Y., Zhuang, F., Xie, R., Zhang, X., Lin, L., He, Q.:
  Personalized transfer of user preferences for cross-domain recommendation.
  In: WSDM. pp. 1507--1515 (2022)

\end{thebibliography}

\end{document}